\documentclass[reprint,prb,superscriptaddress,showpacs]{revtex4-2}%
\usepackage{graphicx}
\usepackage{color}
\usepackage{amsmath}
\usepackage{amsfonts}
\usepackage{amssymb}
\usepackage{framed}
\usepackage{float}
\usepackage[normalem]{ulem}
\usepackage{booktabs}
\usepackage{balance}
\providecommand{\U}[1]{\protect\rule{.1in}{.1in}}
\makeatletter
\renewcommand*{\fnum@figure}{{\normalfont\bfseries \figurename~\thefigure}}
\renewcommand*{\@caption@fignum@sep}{\textbf{ : }}
\makeatother
\usepackage{hyperref}
\hypersetup{
    colorlinks=true,
    linkcolor=blue,
    filecolor=magenta,      
    urlcolor=cyan,
}
\usepackage{textcomp}

\usepackage{lineno}
\begin{document}

\title{ Growth and anisotropic directional dependent transport properties of Altermagnetic CrSb}

% -------------------------------------------------------------------
\author{Resham Babu Regmi}
\email{Corresponding author: rregmi@nd.edu}
\affiliation{Department of Physics and Astronomy, University of Notre Dame, Notre Dame, IN 46556, USA}
\affiliation{Stavropoulos Center for Complex Quantum Matter, University of Notre Dame, Notre Dame, IN 46556, USA}
%\affiliation{Notre Dame Interdisciplinary Materials Science and Engineering Program, University of Notre Dame, Notre Dame, IN 46556, USA}
\author{Hari Bhandari}
\affiliation{Department of Physics and Astronomy, University of Notre Dame, Notre Dame, IN 46556, USA}
\affiliation{Stavropoulos Center for Complex Quantum Matter, University of Notre Dame, Notre Dame, IN 46556, USA}
\affiliation{Department of Physics and Astronomy, George Mason University, Fairfax, VA 22030, USA}
\author{Sk Jamaluddin}
\affiliation{Department of Physics and Astronomy, University of Notre Dame, Notre Dame, IN 46556, USA}
\affiliation{Stavropoulos Center for Complex Quantum Matter, University of Notre Dame, Notre Dame, IN 46556, USA}

\author{Abhijeet Nayak}
\affiliation{Department of Physics and Astronomy, University of Notre Dame, Notre Dame, IN 46556, USA}
\affiliation{Stavropoulos Center for Complex Quantum Matter, University of Notre Dame, Notre Dame, IN 46556, USA}

\author{Ashok Chaudhary}
\affiliation{Department of Physics and Astronomy, University of Notre Dame, Notre Dame, IN 46556, USA}
\affiliation{Stavropoulos Center for Complex Quantum Matter, University of Notre Dame, Notre Dame, IN 46556, USA}

\author{ Mohamed El Gazzah}
\affiliation{Department of Physics and Astronomy, University of Notre Dame, Notre Dame, IN 46556, USA}
\affiliation{Stavropoulos Center for Complex Quantum Matter, University of Notre Dame, Notre Dame, IN 46556, USA}
\author{Alex T. Sheardy}
\affiliation{Notre Dame Integrated Imaging Facility, University of Notre Dame, Notre Dame, IN 46556, USA}

\author{Maksym Zhukovskyi}
\affiliation{Notre Dame Integrated Imaging Facility, University of Notre Dame, Notre Dame, IN 46556, USA}

\author{Allen G. Oliver}
\affiliation{Department of Chemistry and Biochemistry, University of Notre Dame, Notre Dame, IN 46556, USA}

\author{Nirmal J. Ghimire}
\email{Corresponding author: nghimire@nd.edu}
\affiliation{Department of Physics and Astronomy, University of Notre Dame, Notre Dame, IN 46556, USA}
\affiliation{Stavropoulos Center for Complex Quantum Matter, University of Notre Dame, Notre Dame, IN 46556, USA}

% ------------------------------------
\date{\today}
\begin{abstract}

CrSb is a metallic altermagnet exhibiting large momentum-dependent spin splitting and antiferromagnetic ordering well above room temperature. Although its electronic and magnetotransport properties have attracted considerable attention, a systematic comparison of crystals prepared using different growth methods remains limited. Here
we investigate the structural, magnetic, and anisotropic transport properties of CrSb single crystals grown by Sb(self)-flux, Bi flux, and chemical vapor transport (CVT). %Single-crystal XRDX-ray diffraction and electron microscopy confirm the hexagonal NiAs-type structure for all three growth methods, 
Each of these growth reveals distinct crystal morphologies. Magnetization measurements exhibit broadly consistent antiferromagnetic responses, with an additional low-temperature contribution observed in Bi-flux-grown crystals. All samples display metallic resistivity and positive, nonsaturating magnetoresistance up to 9 T. Sb-flux- and CVT-grown crystals exhibit comparable residual resistivity ratios and magnetoresistance, whereas Bi-flux-grown crystals show enhanced residual resistivity and a substantially reduced magnetoresistance, consistent with stronger carrier scattering. Hall measurements reveal a pronounced dependence on the field-current configuration, including nonlinear field dependence and opposite Hall polarities for different crystallographic orientations. These results establish the reproducibility of anisotropic magnetotransport across distinct growth routes while highlighting the sensitivity of carrier scattering to crystal-growth conditions. Our findings provide a comparative framework for selecting CrSb single crystals for orientation-dependent transport and spintronic investigations.

\end{abstract}
\maketitle

\section{Introduction}\label{sec:1}
Altermagnets (AMs) constitute a recently established class of magnetic
materials in which symmetry rules, rather than net magnetization,
dictate the collinear spin texture, distinguishing them from
conventional ferromagnets (FMs) and antiferromagnets (AFMs) while
inheriting useful traits of both for spintronic
functionality~\cite{vsmejkal2022beyond}. %\sout{Each magnetic sublattice is
%internally ferromagnetic, yet adjacent sublattices couple antiparallel,
%yielding a fully compensated antiferromagnetic configuration in real
%space.} 
Unlike conventional collinear AFMs, in which opposite-spin sublattices are related by translation or inversion symmetry, the magnetic sublattices in altermagnets are connected by crystallographic rotations, glide planes or mirror planes.
%The concurrent breaking of time-reversal ($T$), translational
%($t$), and inversion ($P$). 
These symmetries removes the Kramers degeneracy
and generates a momentum-dependent spin splitting whose sign alternates
throughout the Brillouin zone. Because this band structure delivers
spin-split states without invoking spin--orbit coupling while keeping
the macroscopic moment at zero, AMs are attractive for ultrafast,
terahertz-scale spintronic
applications~\cite{vsmejkal2022beyond,mazin2022altermagnetism,vsmejkal2022emerging,feng2022anomalous,krempasky2024altermagnetic,reimers2024direct}.
Altermagnetic signatures have since been documented in a growing list
of compounds from bulk to monolayers, among them are MnTe~\cite{osumi2024observation,krempasky2024altermagnetic,lee2024broken},
CrSb~\cite{reimers2024direct,ding2024large},
CoNb$_4$Se$_8$~\cite{regmi2025altermagnetism,dale2024non,sakhya2025electronic},
$\mathrm{La_2O_3Mn_2Se_2}$ \cite{wei20252},
KV$_2$Se$_2$O~\cite{jiang2025metallic},
RbV$_2$Te$_2$O~\cite{zhang2025crystal},
V$_2$Te$_2$O~\cite{Lee2024} etc.
 
\begin{figure}[ht!]
\centering
\includegraphics[width=1\linewidth]{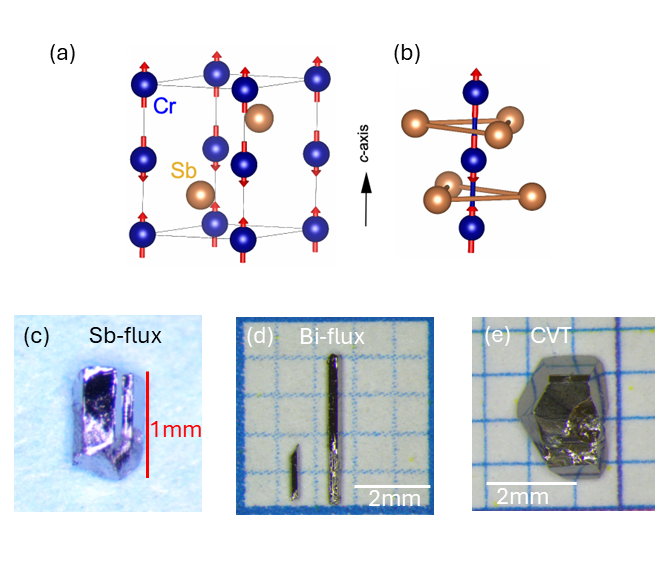}

\caption{\small \textbf{Magnetic structure and optical images.}
(a) Magnetic structure of CrSb with the spin moments aligned along the
$c$-axis. (b) Schematic of the $PT$ symmetry breaking between the Cr
magnetic sublattices. Optical images of CrSb single crystals grown by
(c) Sb-flux, (d) Bi-flux, and (e) CVT.}
\label{figa}
\end{figure}

Among these compounds, CrSb stands out due to its large spin splitting ($\sim$ 1 eV)~\cite{reimers2024direct,zhou2024crystal}.
It adopts the hexagonal NiAs-type structure (space group
$P6_3/mmc$). Cr atoms order antiferromagnetically at relatively high temperature ($\sim$ 700 K) \cite{regus2012investigations} forming an A-type antiferromagnetic arrangement with moments pointing along the crystallographic $c$ -axis as illustrated in Fig.~\ref{figa}(a). As such, CrSb has recently
been studied for a broad range of properties, including
N\'eel-vector-dependent anomalous
transport~\cite{yu2025neel}, conventional charge transport, the Nernst
effect~\cite{li2025large}, spin-dependent
transport~\cite{shi2025spin}, chiral magnon
splitting~\cite{zhang2025chiral,singh2025chiral}, and resonant
inelastic X-ray scattering~\cite{biniskos2025systematic}. Such
measurements have been carried out on both thin films and bulk single
crystals.

%Fuelled by the present momentum in altermagnetism research, CrSb has
%been examined across a broad range of physical properties, including
%N\'eel-vector-dependent anomalous
%transport~\cite{yu2025neel}, conventional charge transport, the Nernst
%effect~\cite{li2025large}, spin-dependent
%transport~\cite{shi2025spin}, chiral magnon
%splitting~\cite{zhang2025chiral,singh2025chiral}, and resonant
%inelastic X-ray scattering~\cite{biniskos2025systematic}. Such
%measurements have been carried out on both thin films and bulk single
%crystals, with the latter generally favoured because they permit
%high-precision probing along well-defined anisotropic crystallographic
%axes.
 
Single-crystal CrSb can be obtained through several routes,
including flux growth using
Sb-flux~\cite{yang2025three,lu2025signature}, Bi-flux and
Sn-flux~\cite{li2025large}, as well as chemical vapor transport (CVT)
employing transport agents such as I$_2$~\cite{li2025topological} or
CrCl$_3$~\cite{singh2025chiral}. The resulting crystals adopt
characteristic morphology, typically needle-like or plate-like, that in
turn dictate which measurement geometries are accessible. Anomalous
Nernst effect studies, for instance, have used needle-shaped crystals
oriented along the $c$-axis, although the limited transverse cross
section of such samples restricts measurements along the remaining
crystallographic directions~\cite{li2025large}. While the magnetic and
transport behavior of CrSb prepared by various synthesis methods has
been reported
extensively~\cite{bai2025nonlinear,urata2024high,peng2025scaling,rai2025direction},
a systematic side-by-side comparison of crystals produced by different
techniques under equivalent conditions is still missing.
 
In this work, we report the growth of single-crystal CrSb by three
complementary methods: a molten-metal flux route using Sb self-flux, a
Bi-flux route, and chemical vapor transport (CVT). The three approaches
yield crystals with distinct sizes, 
and morphological
orientations.
 Structural and compositional analysis was performed by single-crystal
X-ray diffraction (SC-XRD), electron diffraction imaging, and
energy-dispersive X-ray spectroscopy (EDX). We find that the chosen
growth method governs the crystal morphology: Sb-flux growth produces
rod-shaped crystals, Bi-flux growth gives needle-like crystals, and CVT
growth yields plate-like crystals, with the $c$-axis running along the
long dimension for the flux-grown samples and oriented out of plane for
the CVT-grown crystals. Atomic-resolution electron microscopy and
electron diffraction reveal no appreciable differences in lattice
parameters between the samples, beyond minor variations. Magnetic susceptibility measured up to
$400$~K decreases monotonically with no magnetic transitions, as
expected given the high N\'eel temperature of CrSb ($\sim$$700$~K).
Electrical resistivity confirms metallic conduction in every sample,
with the Bi-flux crystals exhibiting a slightly lower residual
resistivity ratio (RRR) compared with the Sb-flux and CVT-grown
crystals. Magnetoresistance is correspondingly large for the Sb-flux
and CVT samples but noticeably smaller for the Bi-flux samples. 

Given the considerable attention CrSb has attracted as a prototypical altermagnet exhibiting exceptionally large momentum-dependent spin splitting, the availability of high-quality single crystals with well-defined crystallographic orientations is crucial for advancing experimental investigations of its unconventional electronic and magnetic properties. Our systematic comparison of three distinct growth methods provides a practical guide to select the most suitable synthesis approach based on the desired crystal morphology, orientation, and transport characteristics.

%As CrSb emerges as a model metallic altermagnet for transport and spintronic investigations, establishing the relationship between crystal growth, structure and the electronic properties is essential. This work provides a systematic comparison of the major crystal-growth techniques used for CrSb and demonstrates that the transport properties relevant to altermagnetic phenomena are largely reproducible across growth methods, with Bi-flux-grown crystals showing somewhat inferior transport properties than the Sb-flux and CVT-grown samples. At the same time, each growth technique yields crystals with distinct morphologies and crystallographic orientations that may be advantageous for specific experimental measurements. These findings provide practical guidance for selecting crystal-growth methods for future studies and enable more reliable comparison of transport data reported by different research groups.
%Taken together, these observations point to superior crystalline quality in the Sb-flux and CVT-grown CrSb single crystals
%%We discuss the results of these characterization techniques which demonstrates the quality of our samples. In addition, we discuss the results of magnetotransport measurements for samples grown via two different approaches  using Physical Property Measurement System (PPMS). 

\section{Materials and Methods}

\begin{table*}[htbp]
  \centering
  \hrule height 1pt
  \caption{\textbf{Crystal data and structure refinement parameters
  for CrSb samples grown by Sb-flux, Bi-flux, and CVT.
  Numbers in parentheses are estimated standard deviations.}}
  \label{Table1}
  \hrule height 1pt
  \vspace{0.3cm}

  \begin{tabular}{lccc}
  \toprule
  \textbf{Compound-CrSb} & \textbf{Sb-flux} &
  \textbf{Bi-flux} & \textbf{CVT} \\
  \midrule
  Crystal system & Hexagonal & Hexagonal & Hexagonal \\
  Space group & $P6_3/mmc$ & $P6_3/mmc$ & $P6_3/mmc$ \\
  Temperature (K) & $120(2)$ & $120(2)$ & 250.00 \\
  Wavelength (\AA) & 0.71073 & 0.71073 & 0.71073 \\
  Z formula units & 2 & 2 & 2 \\
  $\theta_{\min}$ ($^\circ$) & 5.720 & 5.723 & 5.720 \\
  $\theta_{\max}$ ($^\circ$) & 33.147 & 28.031 & 28.140 \\
  Formula weight (g/mol) & 173.75 & 173.75 & 173.75 \\
  $a$, $b$ (\AA) & 4.11680(10) & 4.1148(9) & 4.1171(2) \\
  $c$ (\AA) & 5.4340(4) & 5.4166(15) & 5.4527(3) \\
  Volume (\AA$^3$) & 79.757(7) & 79.42(4) & 80.043(9) \\
  Density (g/cm$^3$) & 7.235 & 7.265 & 7.209 \\
  $\mu$ (Mo K$\alpha$) (mm$^{-1}$) & 23.070 & 23.167 & 22.988 \\
  Goodness-of-fit on $F^2$ & 1.543 & 1.370 & 1.676 \\
  $R(F)$ for $F_o^2 > 2\sigma(F_o^2)$ & 0.0180 & 0.0203 & 0.0114 \\
  $R_w(F_o^2)$ & 0.0374 & 0.0521 & 0.0259 \\
  \bottomrule
\end{tabular}

  \vspace{1em}
  \raggedright
  \textsuperscript{a}
  $R(F) = \sum ||F_o| - |F_c|| / \sum |F_o|$ \\
  \textsuperscript{b}
  $R_w(F^2) =
  \left[\sum w(F_o^2 - F_c^2)^2 /
  \sum w(F_o^2)^2 \right]^{1/2}$

  \vspace{0.3cm}
  \hrule height 1pt
\end{table*}

\begin{figure*}[!ht]
\begin{center}
\includegraphics[width=0.5\linewidth]{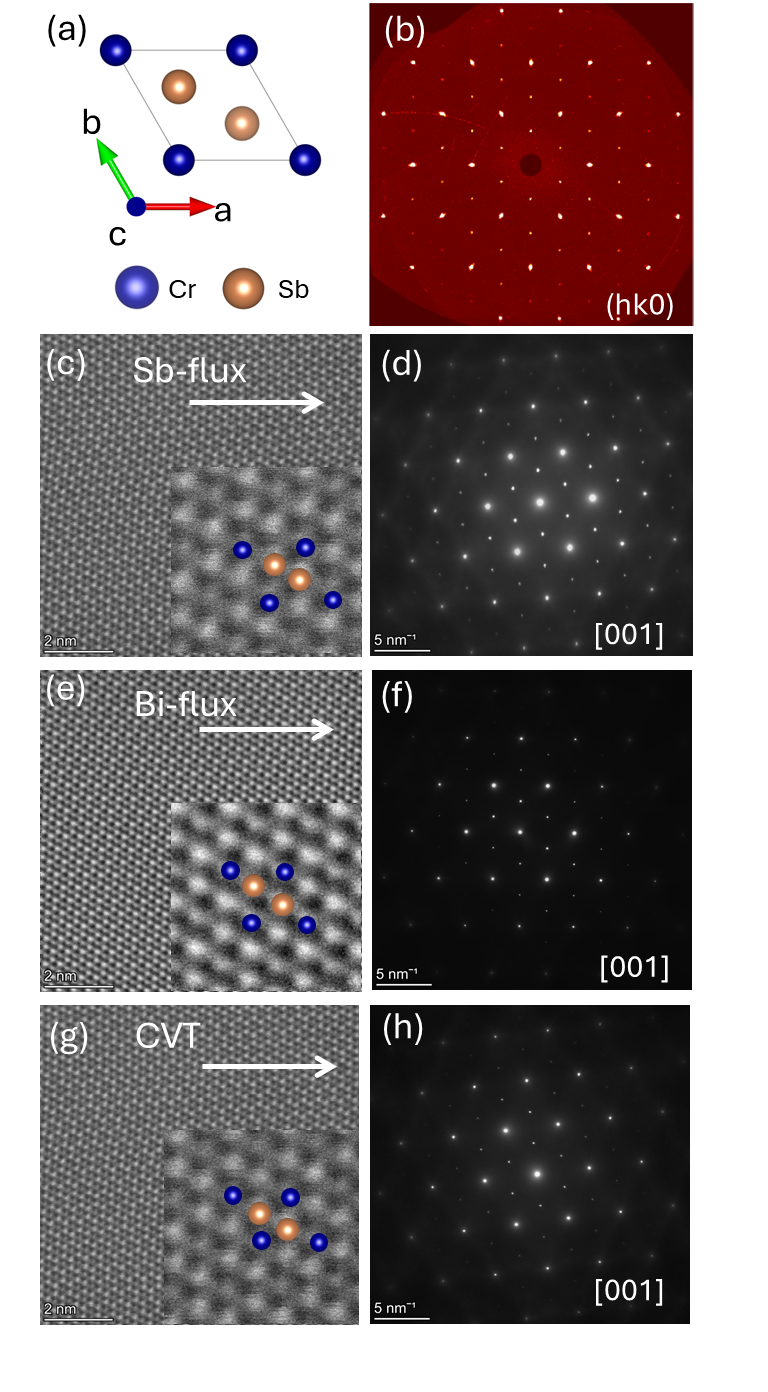}
    \caption{\small \textbf{Crystal structure, single crystal precession and electron microscopy image}  (a) $c$-axis ([001] zone axis) view of the crystal unit cell of CrSb (b) Single Crystal precession image of the measured $hk0$ plane on CVT grown sample(c) HAADF STEM image of Sb-flux CrSb (d) Electron diffraction image of  Sb-flux CrSb (e) HAADF STEM image of Bi-flux CrSb (f) Electron diffraction image of  Bi-flux CrSb (g) HAADF STEM image of CVT CrSb (h) Electron diffraction image of  CVT CrSb. Cr and Sb atoms are displayed as an overlay on top
of atomic resolution image in the inset of panels (c), (e) and (g).}
    \label{crsbbi}
    \end{center}
\end{figure*}

\textbf{Single crystal growth}

\textbf{(i) Sb self-flux method:} Cr powder (Alfa Aesar, 99.99\%) and Sb shots (Thermo Scientific, 99.9999\%) were combined in a 45:55 molar ratio and loaded into a 2~mL crucible, which was then sealed under vacuum in a fused silica ampule. The sealed ampule was heated to 1175~$^{\circ}$C over 10~hours, held at 1175~$^{\circ}$C for 10~hours, and subsequently cooled to 800~$^{\circ}$C over 100~hours. Once the furnace reached 800~$^{\circ}$C, the ampule was centrifuged to decant the molten flux. %Several long needle like  crystals as shown in Fig. \ref{figa}(c) were obtained.
The crystals grown this way were covered with a brown layer, which could be removed simply by using a razor blade to get a shiny single crystal as depicted in Fig. \ref{figa} (c). The crystals obtained by this method had masses of up to 5~mg, with typical dimensions of approximately $2~\mathrm{mm} \times 0.5~\mathrm{mm} \times 0.5~\mathrm{mm}$.

\textbf{(ii) Bi-flux:} Cr powder (Alfa Aesar, 99.99\%), Sb shot (Thermo Scientific, 99.9999\%), and polycrystalline Bi lumps (Alfa Aesar, 99.998\%) were were loaded into a 2 mL crucible in a molar ratio of 1:1:5 successively, which was then sealed under vacuum in a fused silica ampule. A temperature profile similar to that used for the Sb-flux growth was used. The crystals grown by this method were significantly larger along the length, with typical dimensions of approximately $7\,\mathrm{mm} \times 0.2\,\mathrm{mm} \times 0.2\,\mathrm{mm}$. An optical image of a representative Bi-flux-grown crystal is shown in Fig. \ref{figa}(d). 

\textbf{(iii) Chemical Vapor Transport (CVT):} Iodine was used as the transport agent. First, Sb shots (thermo scientific 99.9999\%) was ground into a powder using a mortar and pestle and mixed with chromium powder (Alfa Aesar 99.998\%) in a 1:1 molar ratio. The mixed powder was loaded together with 0.2 g of iodine in a fused silica tube of 14 mm inner diameter. The tube was evacuated and sealed under vacuum. The ampule (10 cm long) was then placed in a two-zone horizontal tube furnace in which the temperature of the hot zone was kept at 850$^{\circ}$C and that of the cold zone was  750$^{\circ}$C for seven days. Several CrSb crystals were formed with a distinct, well-faceted flat plate-like morphology, as shown in Fig. \ref{figa}(e).

\textbf{Single crystal X-ray diffraction:} Structural characterization was performed using single-crystal X-ray diffraction (XRD) at T = 120 K. Data were collected on single crystal grown by each method using a Bruker Quest diffractometer equipped with a Bruker PHOTON III detector, and employing a combination of $\omega$- and $\phi$-scans with a step size of 0.5 degree \cite{apex4}. The diffraction data were corrected for absorption and polarization effects and analyzed to determine the appropriate space group \cite{krause2015}. Structural solution was carried out using dual-space algorithms, followed by routine expansion of the initial model. The final structural refinement was performed using full-matrix least-squares analysis against $F^2$ values for all observed reflections \cite{sheldrick2015shelxt}. The extinction coefficient was also refined. A measured (hk0) zone of a CrSb single crystal is shown in Fig. 2(b). The crystal orientation was confirmed prior to each measurement using a Photonic Science X-ray Laue Crystal Orientation System. 

\textbf{Electron Microscopy:} 
The elemental composition was analyzed on the surface of each of these single crystals using
Magellan 400 - field emission scanning electron microscope (FESEM) with extra high spatial resolution (XHR): 0.6 nm @ 15 kV. Elemental maps were acquired at an accelerating voltage of 15 keV. Lamella of the synthesized materials were prepared for transmission electron microscopy (TEM) using standard lift-out techniques on a Helios G4 (Thermo Fisher Scientific, USA) focused ion beam-scanning electron microscope (FIB-SEM) with a Ga ion source. High angle annular dark field-scanning transmission electron microscopy (HAADF-STEM) and selected area electron diffraction (SAED) were collected on an aberration-corrected Spectra 300 STEM (Thermo Fisher Scientific, USA) operated at 300 kV. Drift-corrected frame integration (DCFI) and an average filter were used to enhance image quality of STEM micrographs.

\textbf{Magnetic measurements:} Magnetic properties were measured using a 7-T Quantum Design Magnetic Property Measurement System (MPMS3) over a temperature range of 1.8 to 400 K. Each of the single crystal samples was mounted on a quartz holder using GE varnish. Magnetization and magnetic susceptibility were measured along both the $ab$-plane and the $c$-axis after confirming the crystal orientation.

\textbf{Electrical transport measurements:}
Electrical resistivity, magnetoresistance, and Hall resistivity were
measured between 1.8 and 300~K in a 9-T Quantum Design DynaCool Physical
Property Measurement System (PPMS). All measurements employed a standard
four-probe configuration, with 25-\textmu m Pt wires serving as electrical
leads. The wires were attached using Epotek H20E silver epoxy cured at
120~$^{\circ}$C, yielding contact resistances below 30~$\Omega$. An
excitation current of 2~mA was used throughout.

Each single crystal was trimmed and polished into a rectangular Hall bar of approximate dimension of $1.00 \times 0.40 \times 0.15$~mm, except for the  Bi-flux-grown crystals for in-plane current measurements. Because of very small diameter of these crystals ($\leq$ 0.2~mm), nanofabrication was required. In this case, a Hall bar device (Figure \ref{device}) was fabricated via Pt deposition using a Ga$^{+}$-based focused ion beam (FIB) .

The longitudinal resistivity, $\rho_{ii}$, was extracted from the measured
longitudinal resistance, $R_{ii}$, through the geometric relation
$\rho_{ii} = R_{ii}A/l$, where $A = td$ is the cross-sectional area ($t$
and $d$ being the sample thickness and width, respectively) and $l$ is the
distance between the voltage contacts. The Hall resistivity was computed as
$\rho_{ij} = R_{ij}t$, where $R_{ij} = V_i/I_j$ is the measured Hall
resistance. In this notation, $V_i$ is the transverse voltage developed
along the $i$-direction for a current $I_j$ applied along the
$j$-direction under a magnetic field along $k$; the mutually orthogonal
indices $i$, $j$, and $k$ correspond to the $x$, $y$, and $z$ axes of the
respective measurement geometries. The magnetoresistance is calculated using relation
$\mathrm{MR} = [\rho_{xx}(B) - \rho_{xx}(0)]/\rho_{xx}(0)$.

To remove artifacts introduced by contact misalignment, the symmetric
(magnetoresistance) and antisymmetric (Hall) components of the measured
data were isolated according to
\begin{equation}
    \rho_{ii} = \frac{\rho_{ii}(+B) + \rho_{ii}(-B)}{2}, \qquad
    \rho_{ij} = \frac{\rho_{ij}(+B) - \rho_{ij}(-B)}{2}.
    \label{eq:symmetrization}
\end{equation}
The field-dependent data were acquired in a four-loop protocol: the
magnetic field $B$ was swept from $+B_{\mathrm{max}}$ to
$-B_{\mathrm{max}}$ and back to $+B_{\mathrm{max}}$, and the
symmetrization and antisymmetrization procedures were applied separately
to the segments spanning $+B_{\mathrm{max}} \rightarrow 0$,
$-B_{\mathrm{max}} \rightarrow 0$, $0 \rightarrow +B_{\mathrm{max}}$, and
$0 \rightarrow -B_{\mathrm{max}}$.

\section{Results and Discussion}\label{sec:4}
\begin{figure*}[!ht]
\begin{center}
\includegraphics[width=1\linewidth]{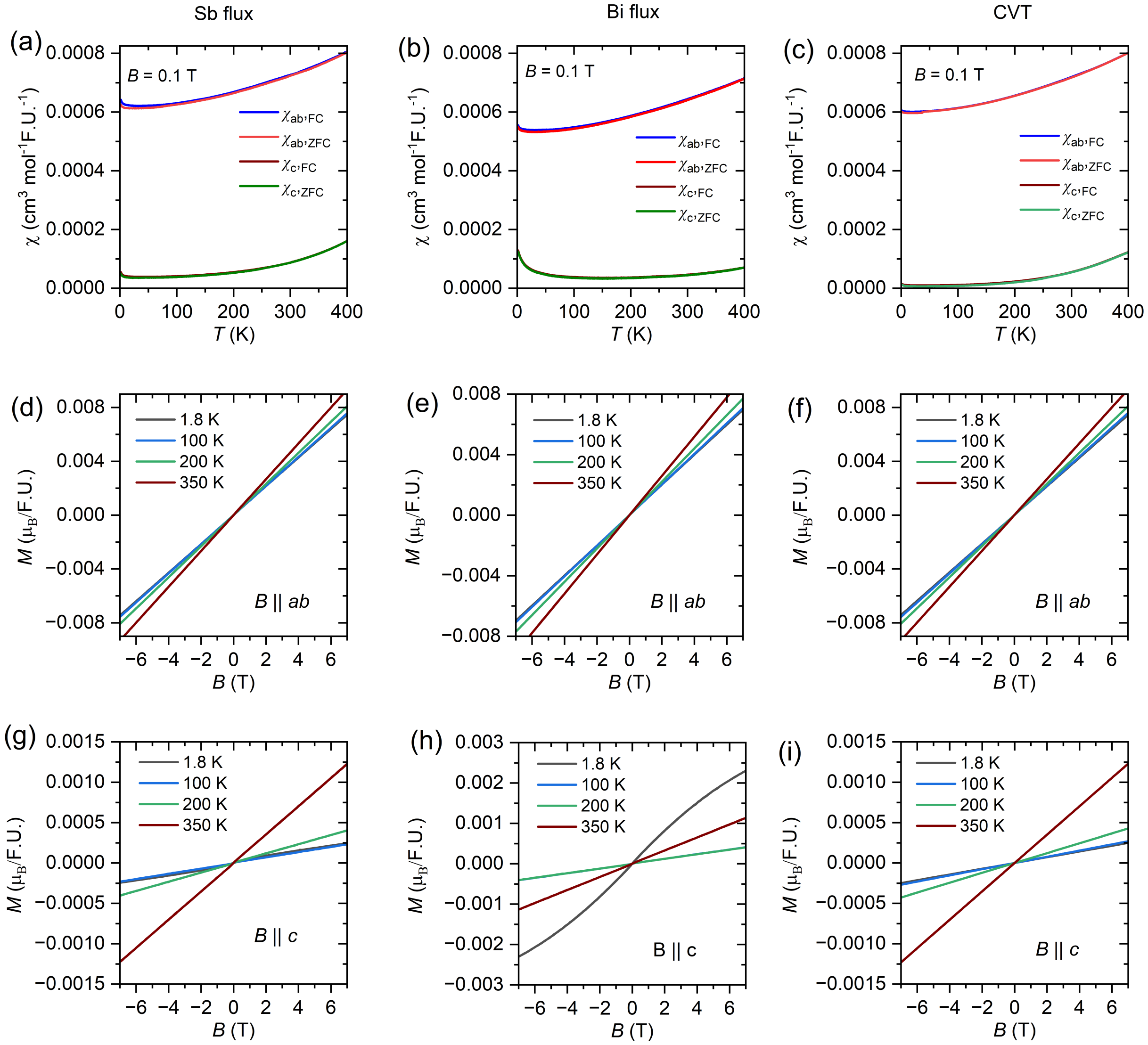}
    \caption{\small \textbf{Magnetic properties of CrSb single
crystals grown by Sb self-flux, Bi flux, and chemical
vapor transport (CVT).}
(a)--(c) Temperature-dependent magnetic susceptibility
$\chi(T)$ for Sb-flux-, Bi-flux-, and CVT-grown
crystals, respectively, measured with
$B\parallel ab$ and $B\parallel c$.
(d)--(f) Field-dependent magnetization $M(B)$
for $B\parallel ab$ for the corresponding
Sb-flux-, Bi-flux-, and CVT-grown crystals.
(g)--(i) Field-dependent magnetization $M(B)$
for $B\parallel c$ for the same respective samples.  }
    \label{MT-MH all}
    \end{center}
\end{figure*}
\subsection{Crystal Structure}

\begin{table}[htbp]
\caption{Atomic coordinates, Wyckoff positions, site occupancies, and
equivalent isotropic displacement parameters (\AA$^{2}$) for Sb-flux grown CrSb.
U(eq) is defined as one third of the trace of the orthogonalized
U$_{ij}$ tensor.}
\label{Table2}
\begin{ruledtabular}
\begin{tabular}{lllllll}
Atom & Wyckoff & Occupancy & $x$ & $y$ & $z$ & U(eq) \\
\colrule
Cr(1) & $2a$ & 1 & 0.00000 & 1.00000 & 0.50000 & 0.003(1) \\
Sb(1) & $2c$ & 1 & 0.33333 & 0.66667 & 0.75000 & 0.001(1)    \\
\end{tabular}
\end{ruledtabular}
\end{table}

Single-crystal X-ray diffraction (SC-XRD) confirms that
CrSb crystals grown by all three methods have the same hexagonal NiAs-type structure. A precession image of a representative single crystal data measured in CVT grown sample is presented in Fig. \ref{crsbbi}b. The details of refinement from SC-XRD are provided in Table \ref{Table1}-\ref{table 11} (excluding \ref{table 3}). The refined lattice parameters are $a=b=4.11680(10)$~\AA{} and
$c=5.4340(4)$~\AA{} for Sb-flux-grown crystals,
$a=b=4.1148(9)$~\AA{} and $c=5.4166(15)$~\AA{}
for Bi-flux-grown crystals, and
$a=b=4.1171(2)$~\AA{} and $c=5.4527(3)$~\AA{}
for CVT-grown crystals. This shows that, although these crystals have the same structure, there is small variations in the lattice parameters. 
The in-plane lattice parameter varies by less than 0.06\%, whereas the out-of-plane lattice parameter exhibits a comparatively larger variation of approximately 0.67\%. Notably, CVT-grown crystals exhibit the largest \(c\)-axis lattice parameter, followed by Sb-flux- and Bi-flux-grown crystals. These observations suggest that the crystallographic \(c\)-axis is particularly sensitive to growth conditions, potentially reflecting subtle differences in stoichiometry, defect concentrations, or residual strain. Given the sensitivity of electronic structure to interatomic distances and orbital hybridization, these growth-dependent structural variations may be relevant to future investigations of the altermagnetic properties of CrSb.

%All three samples retain the same crystallographic structure, although variations in the lattice parameters are observed among the different growth methods. The complete crystallographic refinement parameters are summarized in Table~I, while the atomic coordinates, site occupancies, and equivalent isotropic displacement parameters are provided in Table~II for the Sb-flux-grown sample and in Appendix Tables~III and IV for the Bi-flux- and CVT-grown samples, respectively \cite{snow1952neutron, willis1953crystal,takei1963magnetic,reimers1982magnetic}.

To further investigate the microscopic structural characteristics of CrSb single crystals and assess potential variations arising from different growth conditions, we performed transmission electron microscopy (TEM) and selected-area electron diffraction (SAED) measurements on crystals grown using all three methods. The samples were oriented along the [001] zone axis, with the corresponding crystallographic structure illustrated in Fig.~\ref{crsbbi}(a). The high-angle annular dark-field (HAADF) images [Figs.~\ref{crsbbi}(c), (e), and (g)] reveal well-ordered atomic arrangements consistent with the expected crystal structure of CrSb. Furthermore, the corresponding electron diffraction patterns [Figs.~\ref{crsbbi}(d), (f), and (h)] exhibit distinct periodic features, including additional diffraction features resembling those observed in the precession X-ray diffraction image [Fig.~\ref{crsbbi}(b)]. Analysis of the SAED patterns yields in-plane lattice parameters of $a = 4.13(1)$, $4.07(3)$, and $4.14(1)$~\AA{} for the Sb-flux, Bi-flux, and CVT-grown crystals, respectively with the number in the parentheses being the estimated standard uncertainties. These values are in good agreement with those obtained from SC-XRD, confirming the overall structural consistency of CrSb crystals grown using different techniques. Although SC-XRD reveals subtle growth-dependent variations in the lattice parameters, the uncertainties associated with SAED measurements preclude resolving such small differences. Moreover, the [001] orientation of the TEM lamellae restricts the diffraction analysis to the basal plane, preventing an independent determination of the out-of-plane lattice parameter, $c$. Together, these complementary measurements confirm the crystallographic consistency of CrSb across different growth methods, while revealing subtle structural variations that can be resolved by SC-XRD but not by SAED.

\subsection{Magnetic Susceptibility and Magnetization}

\begin{figure*}[!ht]
\begin{center}
\includegraphics[width=0.9\linewidth]{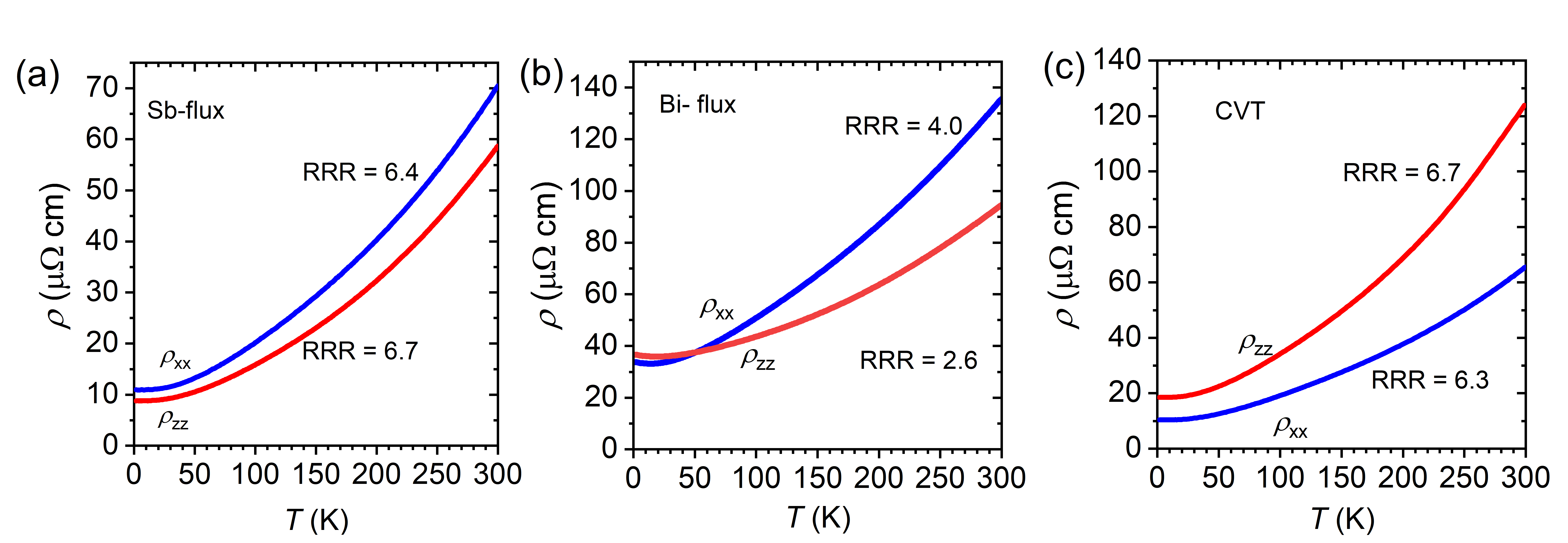}
    \caption{\small \textbf{Electrical transport measurements of CrSb samples}:  Resistivity as a function of temperature (a)  for Sb-flux  (b) Bi-flux (c) CVT  along $I\parallel c$  and $I\parallel a$ orientations. }
    
    \label{merged RT}
    \end{center}
\end{figure*}

The magnetic properties of CrSb single crystals grown using
the three different methods are presented in
Fig.~\ref{MT-MH all}; where (a--c) 
show the temperature-dependent magnetic susceptibility,
$\chi(T)$ of Sb-flux, Bi-flux and CVT grown samples, respectively. These data are measured between 1.8 and 400~K under an applied
magnetic field of 0.1~T oriented either parallel to the
crystallographic $c$ axis ($\chi_c$) or within the basal
plane ($\chi_{ab}$). No magnetic phase transition is
observed over the investigated temperature range,
consistent with the high N\'eel temperature of CrSb
($T_{\mathrm{N}}\sim700$~K) established by previous
neutron diffraction studies
\cite{snow1952neutron,takei1963magnetic}.
For all three growth methods, the magnetic susceptibility
generally decreases upon cooling, consistent with the
characteristic antiferromagnetic response previously
reported for CrSb \cite{bai2025nonlinear}.

A pronounced magnetic anisotropy is observed in all
samples, with $\chi_{ab}>\chi_c$ throughout the
measured temperature range. This behavior is consistent
with the uniaxial magnetic anisotropy of CrSb, whose
collinear antiferromagnetic structure features Cr moments
aligned along the crystallographic $c$ axis, as
established by neutron diffraction experiments
\cite{yuan2020magnetic}.

Despite the different growth conditions, all three
samples exhibit qualitatively similar magnetic
susceptibility, indicating that the overall magnetic
response is largely independent of the growth method.
However, slight difference emerges in the
low-temperature $c$-axis susceptibility of the
Bi-flux-grown crystal, which exhibits a comparatively small upturn
that is absent in the Sb-flux- and CVT-grown samples.
This weak upturn may arise from a small extrinsic
magnetic contribution, potentially associated with
growth-related defects, dilute magnetic impurities,
or local variations in the magnetic environment.

The field-dependent magnetization, $M(B)$, of CrSb single
crystals grown using the three different methods is shown
in Figs.~\ref{MT-MH all}(d)--(i). Measurements were
performed at selected temperatures between 1.8 and 350~K,
with magnetic fields up to $\pm 7$~T applied either parallel
to the crystallographic $c$ axis ($B\parallel c$) or within
the basal plane ($B\parallel ab$).

For the Sb-flux- and CVT-grown crystals, the magnetization
exhibits an approximately linear field dependence in both
orientations over the investigated temperature range, with
no evidence of field-induced magnetic transitions up to
7~T. The Bi-flux-grown crystal exhibits a similarly linear
response over most of the measured field and temperature
range. However, a weak deviation from linearity emerges
at 1.8~K for $B\parallel c$, consistent with a slight
low-temperature upturn observed in the corresponding
magnetic susceptibility. This deviation may reflect a
small extrinsic magnetic contribution, potentially
associated with growth-related defects or dilute magnetic
impurities, although its microscopic origin remains
unclear.

Taken together, the temperature- and field-dependent
magnetization measurements demonstrate that the magnetic
response of CrSb is largely consistent across the three
growth methods, with only subtle low-temperature
differences observed in the Bi-flux-grown crystal.

%%Magnetization ($MB$) measurements is presented in Figs. \ref{MT-MH all}. Fig \ref{MT-MH all} all (d),(e),(f) provide further insight into the nature of magnetic ordering in this material. Figure \ref{MT-MH all}(c) shows the $MB$ data measured with $B$ between $\pm$ 7 T applied along the $c$-axis ($MB_{c}$) and $ab$-plane ($MB_{ab}$) at various temperatures ranging from 1.8 K-350 K. For all the samples, the curve $MB_{ab}$ and $MB_{c}$ exhibits linear behavior, a characteristic of an antiferromagnetic behaviour. Only $MB_{c}$ for Bi-flux samples has slight non-linearity at 1.8 K most likely due to the reason mentioned above. This identifies by magnetic susceptibility measurement Sb-flux growth sample and CVT grown sample as having identical property along field applied in plane and out of plane directions while with Bi-flux sample having slight variation at low temperature in $B$ $||$ $c$ direction. 

\subsection{Resistivity, Magnetoresistance and Hall}
\begin{figure*}[!ht]
\begin{center}
\includegraphics[width=0.9\linewidth]{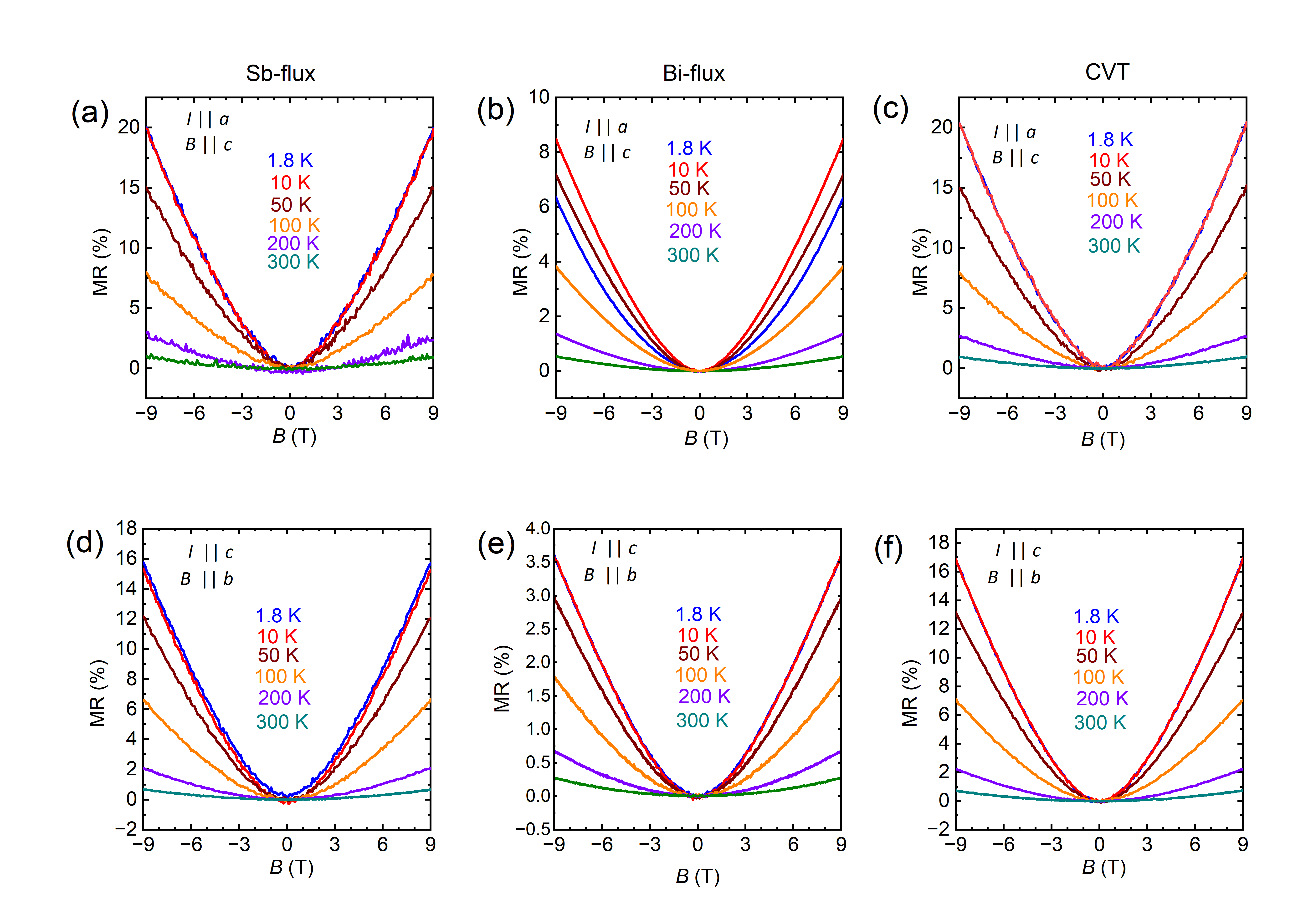}
    \caption{\small \textbf{Magnetotransport measurements of CrSb samples:}   Magnetoresistance of CrSb along  $I\parallel a$ and $B\parallel c$ for samples grown by (a)Sb-flux (b) Bi-flux (c) CVT. Magnetoresistance of CrSb along  $I\parallel c$ and $B\parallel b$ for samples grown by (d)Sb-flux (e) Bi-flux (f) CVT.}
    \label{crsbself}
    \end{center}
\end{figure*}

Figure~\ref{merged RT} presents the temperature-dependent
electrical resistivity of CrSb single crystals grown using
Sb-flux, Bi-flux, and chemical vapor transport (CVT).
Measurements were performed with the electrical current
applied along the crystallographic $c$ and $a$ axes,
corresponding to $\rho_{zz}$ (red) and $\rho_{xx}$ (blue),
respectively. All three samples exhibit metallic behavior,
characterized by a monotonic decrease in resistivity upon
cooling. No pronounced anomalies are observed between
1.8 and 300~K, consistent with the absence of magnetic
phase transitions within this temperature range, as
expected from the high temperature antiferromagnetic ordering of CrSb ($T_{\mathrm{N}}\sim700$~K)
\cite{haraldsen1946system}.

To assess the influence of growth conditions on charge
transport, we calculated the residual resistivity ratio,
defined as $\mathrm{RRR}=\rho(300~\mathrm{K})/
\rho(1.8~\mathrm{K})$. The Sb-flux- and CVT-grown
crystals exhibit comparable RRR values of approximately
6.7 and 6.4 for $\rho_{zz}$ ($I\parallel c$) and $\rho_{xx}$ ($I\parallel a$),
respectively, suggesting similar temperature-dependent
scattering characteristics. In contrast, the Bi-flux-grown
crystals exhibit higher residual resistivity and reduced
RRR values of 4.0 and 2.6 for $\rho_{xx}$ and
$\rho_{zz}$, respectively. These differences suggest
enhanced residual carrier scattering in the Bi-flux-grown
samples, potentially associated with growth-dependent
disorder or compositional variations. Contributions from
Ga implantation during focused ion beam (FIB)
nanofabrication may also influence the measured
resistivity, particularly in the microfabricated device prepared for measuring $\rho_{xx}$ of Bi-flux sample (refer to Figure \ref{device}).

Interestingly, the resistivity exhibits a distinct
growth-dependent anisotropy. In both Sb-flux- and
Bi-flux-grown crystals, the in-plane resistivity
$\rho_{xx}$ generally exceeds the out-of-plane
resistivity $\rho_{zz}$ over most of the measured
temperature range, although a reversal occurs below
approximately 50~K in the Bi-flux-grown sample.
In contrast, CVT-grown crystals exhibit the opposite
anisotropy, with $\rho_{zz}>\rho_{xx}$ throughout the
investigated temperature range. This contrasting
behavior suggests that the anisotropic charge transport
in CrSb may be sensitive to growth conditions.
Because electrical resistivity reflects both the
anisotropy of the electronic band structure and
carrier scattering processes, the observed differences
could arise from growth-dependent variations in
electronic structure, defect concentrations, or
scattering rates. In particular, subtle modifications
of the Fermi surface and associated carrier velocities
may influence the relative magnitudes of $\rho_{xx}$
and $\rho_{zz}$. Further investigations are required
to establish the microscopic origin of this
growth-dependent transport anisotropy.

\begin{figure*}[!ht]
\begin{center}
\includegraphics[width=1\linewidth]{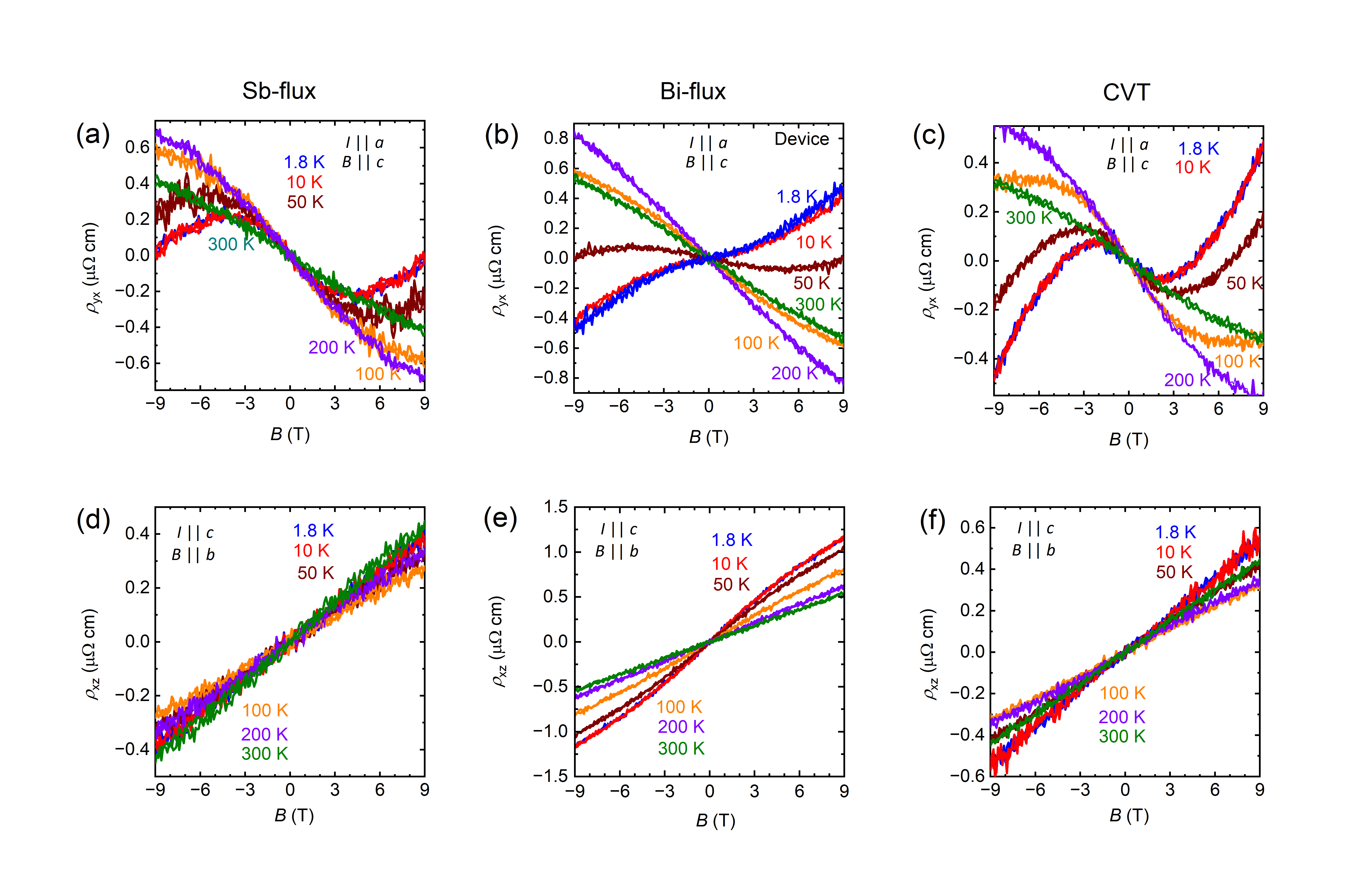}
    \caption{\small \textbf{Hall resistivity measurements of CrSb samples} Hall resistivity  measurements  along $I\parallel a$ and $B\parallel c$ for (a)  Sb-flux CrSb, (b) Bi-flux (c) CVT. Hall resistivity  measurements of CrSb along $I\parallel c$ and $B\parallel b$ for (d) Sb flux, (e) Bi-flux (f) CVT.}
    \label{Hall}
    \end{center}
\end{figure*}

Magnetoresistance of each of these samples is presented
in Figs.~\ref{crsbself}(a)--(f) for two measurement
configurations: $I\parallel a$ (for simplicity $a$ is used for [100]), $B\parallel c$  and
$I\parallel c$ (for simplicity $b$ is used for [001]), $B\parallel b$ (for simplicity $b$ is used for [120]). All three growth types
exhibit positive, nonsaturating magnetoresistance up to
9~T, with an approximately quadratic field dependence
at low fields. This behavior is consistent with an
orbital contribution to magnetotransport and is
compatible with the multicarrier conduction previously
established in CrSb \cite{urata2024high}.

At 1.8~K and 9~T, the MR reaches approximately
16\% for the Sb-flux- and CVT-grown crystals in the
$I\parallel c$, $B\parallel  b$ configuration,
while values approaching 20\% are observed for
$I\parallel a$, $B\parallel c$. These magnitudes
are comparable to previous experimental reports
\cite{bai2025nonlinear,urata2024high} and are
consistent with the magnetotransport behavior
analyzed theoretically in Ref.~\cite{peng2025scaling}.
The MR decreases rapidly with increasing temperature
and becomes relatively small above 200~K, consistent
with the enhanced carrier scattering expected at
elevated temperatures. In contrast, the Bi-flux-grown crystals exhibit
substantially smaller MR values of approximately
8\% for $I\parallel a$, $B\parallel c$ and
3.5\% for $I\parallel c$, $B\parallel b$
at 1.8~K and 9~T. The reduced MR, together with
the lower RRR observed
in these crystals, suggests enhanced carrier
scattering relative to the Sb-flux- and CVT-grown
samples. Thus, while the qualitative field and
temperature dependences of MR remain similar
across the different growth methods, its magnitude
is sensitive to growth-dependent electronic
scattering processes.

\begin{table}[htbp]
    \centering
    \caption{Effective Hall carrier densities of CrSb
    single crystals grown by different methods at
    $T=300$~K. The values are estimated using the
    single-band approximation
    $n_{\mathrm{H}}^{\mathrm{eff}}=1/(eR_{\mathrm{H}})$.}
    \label{table 3}
    \begin{ruledtabular}
    \begin{tabular}{lcc}
        Growth method &
        \multicolumn{2}{c}{
        $n_{\mathrm{H}}^{\mathrm{eff}}$
        ($10^{22}$~cm$^{-3}$)} \\
        \cline{2-3}
        & $\rho_{xz}$ &
          $\rho_{yx}$ \\
        \hline
        Sb-flux &  1.344 & -1.330 \\
        Bi-flux &  1.032 & -1.043 \\
        CVT     &  1.277 & -1.783 \\
    \end{tabular}
    \end{ruledtabular}
\end{table}

The magnetic-field dependence of the Hall resistivity
for CrSb single crystals grown using the three different
methods is presented in Figs.~\ref{Hall}(a)--(f), for
the same two measurement configurations employed in
the magnetoresistance measurements:
$I\parallel a$, $B\parallel c$ ($\rho_{yx}$),
and $I\parallel c$, $B\parallel b$ ($\rho_{xz}$).
For all three growth methods, $\rho_{yx}$ exhibits a
pronounced nonlinear field dependence below approximately
200~K, consistent with multicarrier transport.
In contrast, $\rho_{xz}$ remains predominantly linear
with a positive Hall coefficient, suggesting a
hole-dominated Hall response in this configuration.
The degree of nonlinearity in $\rho_{yx}$ varies
among crystals grown using different methods,
with CVT-grown samples exhibiting the most pronounced
nonlinear response and Bi-flux-grown samples showing
the weakest.

Interestingly, we also observe sample-to-sample
variations in the transport properties of Sb-flux-grown
crystals. While most samples exhibit behavior similar
to that presented in Figs.~\ref{merged RT},
\ref{crsbself}, and \ref{MT-MH all}, a subset of
Sb-flux-grown crystals displays a substantially more
pronounced nonlinear Hall response, accompanied by
a larger magnetoresistance ($\sim90\%$) and a higher
residual resistivity ratio ($\mathrm{RRR}\sim10.8$),
as shown in Fig. \ref{TransportSb} of the Appendix. These observations
suggest that the electronic transport properties of
CrSb are sensitive not only to the growth method but
also to sample-to-sample variations within a given
growth technique, potentially reflecting differences
in defect concentrations, carrier mobilities, or
relative contributions from multiple carrier bands.

The contrasting Hall responses in the two
configurations are consistent with the
direction-dependent conduction polarity
previously reported in CrSb
\cite{rai2025direction}. This behavior has
been associated with the goniopolar topology
of its Fermi surface, which gives rise to
anisotropic, mobility-weighted contributions
from electron- and hole-like carriers.
Consequently, the in-plane Hall response is
predominantly electron-like, whereas the
out-of-plane response is predominantly
hole-like \cite{rai2025direction}. With
increasing temperature, the Hall resistivity
becomes progressively more linear in field,
suggesting that the transport response can
be approximated more closely by an effective
single-carrier description at higher temperatures.

To provide a quantitative comparison among
the different growth methods, the effective
Hall carrier density at 300~K was estimated
using the single-band expression $n_{\mathrm{H}}^{\mathrm{eff}}
    = \frac{1}{eR_{\mathrm{H}}}$, where $e$ is the elementary charge and
$R_{\mathrm{H}}$ is the Hall coefficient
obtained from the slope of the Hall
resistivity. The resulting effective Hall
carrier densities for both measurement
configurations are summarized in Table \ref{table 3}.
Because CrSb is a multiband metal, these
values represent effective single-band
estimates rather than the actual electron
and hole concentrations.

Although crystals obtained by the three
growth methods exhibit broadly similar
crystallographic structures, quantitative
differences are observed in their electrical
transport properties. Nevertheless, the
characteristic directional Hall response
is reproduced across all three growth
methods, despite differences in residual
resistivity and magnetoresistance magnitude.
This reproducibility suggests that the
direction-dependent Hall polarity is
primarily associated with the intrinsic
anisotropic electronic structure of CrSb,
whereas growth-dependent scattering
influences the magnitude and detailed
field dependence of the transport response.

\section{Conclusion}

We have systematically investigated the structural, magnetic, and anisotropic magnetotransport properties of CrSb single crystals grown using Sb self-flux, Bi flux, and chemical vapor transport. All three growth methods yield crystals with the hexagonal NiAs-type structure and broadly consistent magnetic responses, while producing distinct crystal morphologies and differences in their electrical transport properties.

Sb-flux- and CVT-grown crystals exhibit comparable residual resistivity ratios and positive, nonsaturating magnetoresistance, whereas Bi-flux-grown crystals display higher residual resistivity and substantially smaller magnetoresistance. These differences are consistent with enhanced carrier scattering in the Bi-flux-grown samples. Despite these variations, the direction-dependent Hall response is qualitatively reproduced across all three growth methods, including opposite Hall polarities for different field-current configurations and nonlinear field dependence at low temperatures. The reproducibility of these features suggests that the directional Hall response is closely associated with the intrinsic anisotropic electronic structure of CrSb.

Our results distinguish the growth-sensitive and comparatively robust features of charge transport in CrSb and establish a framework for selecting crystals with suitable morphology and transport characteristics for future investigations of altermagnetic electronic phenomena.

\begin{acknowledgments}
 N.J.G and R.B.R. acknowledges support from Army Research Office
under Cooperative Agreement Number W911NF-26-2-A187. R.B.R. also acknowledges support from MSE-ND for innovation fellowship.
A.O. acknowledges funding for instrumentation of
Single Crystal X-ray Diffraction by NSF MRI Award CHE-2214606. R.B.R. also acknowledges Tatyana Orlova for the FIB sample preparation and the NDIF facility at Notre Dame for the FIB and TEM imaging.  
\end{acknowledgments}

% =====================================================
% REFERENCES
% =====================================================

\bibliography{references.bib}

% =====================================================
% FULL-WIDTH SINGLE-COLUMN APPENDIX
% =====================================================
%\newpage

\onecolumngrid
\clearpage
\appendix
\section{Appendix}

% =====================================================
% Sb-FLUX: ANISOTROPIC DISPLACEMENT PARAMETERS
% =====================================================

\begin{table}[!htbp]
    \centering
    \caption{\textbf{Anisotropic displacement parameters
    ($\mathrm{\AA}^2 \times 10^3$) for Sb-flux-grown CrSb.}
    The anisotropic displacement factor exponent takes
    the form
    $-2\pi^2[h^2a^{*2}U_{11}+\cdots+
    2hka^*b^*U_{12}]$.
    Numbers in parentheses represent estimated
    standard deviations.}
    \label{tab:adp_sb_crsb}

    \begin{tabular}{lcccccc}
        \toprule
        Atom & $U_{11}$ & $U_{22}$ & $U_{33}$
             & $U_{23}$ & $U_{13}$ & $U_{12}$ \\
        \midrule
        Sb(1) & 3.1(4) & 3.1(4) & 2.4(4)
              & 0 & 0 & 1.6(2) \\
        Cr(1) & 4.2(4) & 4.2(4) & 2.2(6)
              & 0 & 0 & 2.1(2) \\
        \bottomrule
    \end{tabular}
\end{table}

% =====================================================
% Sb-FLUX: BOND LENGTHS
% =====================================================

\begin{table}[!htbp]
    \centering
    \caption{\textbf{Selected bond lengths (\AA)
    for Sb-flux-grown CrSb.}
    Numbers in parentheses represent estimated
    standard deviations.}
    \label{tab:bond_lengths_sb_crsb}

    \begin{tabular}{lclc}
        \toprule
        Atom pair & Distance (\AA) &
        Atom pair & Distance (\AA) \\
        \midrule
        Sb(1)--Cr(1)\#1 & 2.73768(7) &
        Sb(1)--Cr(1)    & 2.73768(7) \\
        Sb(1)--Cr(1)\#2 & 2.73768(6) &
        Sb(1)--Cr(1)\#3 & 2.73768(7) \\
        Sb(1)--Cr(1)\#4 & 2.73768(7) &
        Sb(1)--Cr(1)\#5 & 2.73768(7) \\
        Cr(1)--Cr(1)\#1 & 2.7170(2) &
        Cr(1)--Cr(1)\#6 & 2.7170(2) \\
        \bottomrule
    \end{tabular}

    \vspace{0.12cm}

    \begin{minipage}{0.85\textwidth}
    \centering
    \footnotesize
    \textit{Symmetry transformations used to generate
    equivalent atoms:}

    \vspace{0.06cm}

    \begin{tabular}{@{}ll@{}}
        \#1: $-x,-y+2,z+\frac{1}{2}$ &
        \#2: $-x+1,-y+2,z+\frac{1}{2}$ \\
        \#3: $x,y-1,z$ &
        \#4: $x+1,y,z$ \\
        \#5: $-x,-y+1,z+\frac{1}{2}$ &
        \#6: $-x,-y+2,z-\frac{1}{2}$ \\
    \end{tabular}
    \end{minipage}
\end{table}

% =====================================================
% Bi-FLUX: ATOMIC COORDINATES
% =====================================================

\begin{table}[!htbp]
    \centering
    \caption{Atomic coordinates, Wyckoff positions,
    site occupancies, and equivalent isotropic
    displacement parameters (\AA$^{2}$) for Bi-flux
    grown CrSb. U(eq) is defined as one third of
    the trace of the orthogonalized U$_{ij}$ tensor.}
    \label{Table3}

    \begin{ruledtabular}
    \begin{tabular}{lllllll}
        Atom & Wyckoff & Occupancy &
        $x$ & $y$ & $z$ & U(eq) \\
        \colrule
        Cr(1) & $2a$ & 1 & 0.00000 & 1.00000 &
        0.50000 & 0.005(1) \\
        Sb(1) & $2c$ & 1 & 0.33333 & 0.66667 &
        0.25000 & 0.004(1) \\
    \end{tabular}
    \end{ruledtabular}
\end{table}

% =====================================================
% Bi-FLUX: ANISOTROPIC DISPLACEMENT PARAMETERS
% =====================================================

\begin{table}[!htbp]
    \centering
    \caption{\textbf{Anisotropic displacement parameters
    ($\mathrm{\AA}^2 \times 10^3$) for Bi-flux-grown CrSb.}
    The anisotropic displacement factor exponent takes
    the form
    $-2\pi^2[h^2a^{*2}U_{11}+\cdots+
    2hka^*b^*U_{12}]$.
    Numbers in parentheses represent estimated
    standard deviations.}
    \label{tab:adp_bi_crsb}

    \begin{tabular}{lcccccc}
        \toprule
        Atom & $U_{11}$ & $U_{22}$ & $U_{33}$
             & $U_{23}$ & $U_{13}$ & $U_{12}$ \\
        \midrule
        Sb(1) & 5.5(10) & 5.5(10) & 1.8(10)
              & 0 & 0 & 2.7(5) \\
        Cr(1) & 6.7(9) & 6.7(9) & 2.3(12)
              & 0 & 0 & 3.3(5) \\
        \bottomrule
    \end{tabular}
\end{table}

% =====================================================
% Bi-FLUX: BOND LENGTHS
% =====================================================

\begin{table}[!htbp]
    \centering
    \caption{\textbf{Selected bond lengths (\AA) for
    Bi-flux-grown CrSb.} Numbers in parentheses
    represent estimated standard deviations.}
    \label{tab:bond_lengths_bi_crsb}

    \begin{tabular}{lclc}
        \toprule
        Atom pair & Distance (\AA) &
        Atom pair & Distance (\AA) \\
        \midrule
        Sb(1)--Cr(1)\#1 & 2.7345(5) &
        Sb(1)--Cr(1)    & 2.7345(5) \\
        Sb(1)--Cr(1)\#2 & 2.7345(5) &
        Sb(1)--Cr(1)\#3 & 2.7345(5) \\
        Sb(1)--Cr(1)\#4 & 2.7345(5) &
        Sb(1)--Cr(1)\#5 & 2.7345(5) \\
        Cr(1)--Cr(1)\#6 & 2.7083(8) &
        Cr(1)--Cr(1)\#1 & 2.7083(8) \\
        \bottomrule
    \end{tabular}

    \vspace{0.12cm}

    \begin{minipage}{0.85\textwidth}
    \centering
    \footnotesize
    \textit{Symmetry transformations used to generate
    equivalent atoms:}

    \vspace{0.06cm}

    \begin{tabular}{@{}ll@{}}
        \#1: $-x,-y+2,z-\frac{1}{2}$ &
        \#2: $-x,-y+1,z-\frac{1}{2}$ \\
        \#3: $x+1,y,z$ &
        \#4: $-x+1,-y+2,z-\frac{1}{2}$ \\
        \#5: $x,y-1,z$ &
        \#6: $-x,-y+2,z+\frac{1}{2}$ \\
    \end{tabular}
    \end{minipage}
\end{table}

% =====================================================
% CVT: ATOMIC COORDINATES
% =====================================================

\begin{table}[!htbp]
    \centering
    \caption{Atomic coordinates, Wyckoff positions,
    site occupancies, and equivalent isotropic
    displacement parameters (\AA$^{2}$) for CVT
    grown CrSb. $U_{eq}$ is defined as one third
    of the trace of the orthogonalised
    $U_{ij}$ tensor.}
    \label{Table4}

    \begin{ruledtabular}
    \begin{tabular}{lllllll}
        Atom & Wyckoff & Occupancy &
        $x$ & $y$ & $z$ & $U_{eq}$ \\
        \colrule
        Cr(1) & $2a$ & 1 & 0.00000 & 1.00000 &
        0.50000 & 0.0061(4) \\
        Sb(1) & $2c$ & 1 & 0.33333 & 0.66667 &
        0.25000 & 0.0062(4) \\

    \end{tabular}
    \end{ruledtabular}
\end{table}

% =====================================================
% CVT: ANISOTROPIC DISPLACEMENT PARAMETERS
% =====================================================

\begin{table}[!htbp]
    \centering
    \caption{\textbf{Anisotropic displacement parameters
    ($\mathrm{\AA}^2 \times 10^3$) for CVT-grown CrSb.}
    The anisotropic displacement factor exponent takes
    the form
    $-2\pi^2[h^2a^{*2}U_{11}+\cdots+
    2hka^*b^*U_{12}]$.
    Numbers in parentheses represent estimated
    standard deviations.}
    \label{tab:adp_cvt_crsb}

    \begin{tabular}{lcccccc}
        \toprule
        Atom & $U_{11}$ & $U_{22}$ & $U_{33}$
             & $U_{23}$ & $U_{13}$ & $U_{12}$ \\
        \midrule
        Sb(1) & 5.8(4) & 5.8(4) & 7.0(4)
              & 0 & 0 & 2.9(2) \\
        Cr(1) & 6.6(4) & 6.6(4) & 5.2(6)
              & 0 & 0 & 3.3(2) \\
        \bottomrule
    \end{tabular}
\end{table}

% =====================================================
% CVT: BOND LENGTHS
% =====================================================

\begin{table}[!htbp]
    \centering
    \caption{\textbf{Selected bond lengths (\AA)
    for CVT-grown CrSb.}
    Numbers in parentheses represent estimated
    standard deviations.}
    \label{table 11}

    \begin{tabular}{lclc}
        \toprule
        Atom pair & Distance (\AA) &
        Atom pair & Distance (\AA) \\
        \midrule
        Sb(1)--Cr(1)\#1 & 2.74015(10) &
        Sb(1)--Cr(1)    & 2.74015(10) \\
        Sb(1)--Cr(1)\#2 & 2.74015(10) &
        Sb(1)--Cr(1)\#3 & 2.74015(10) \\
        Sb(1)--Cr(1)\#4 & 2.74015(10) &
        Sb(1)--Cr(1)\#5 & 2.74015(11) \\
        Cr(1)--Cr(1)\#1 & 2.72635(15) &
        Cr(1)--Cr(1)\#6 & 2.72635(15) \\
        \bottomrule
    \end{tabular}

    \vspace{0.12cm}

    \begin{minipage}{0.85\textwidth}
    \centering
    \footnotesize
    \textit{Symmetry transformations used to generate
    equivalent atoms:}

    \vspace{0.06cm}

     \begin{tabular}{@{}ll@{}}
        \#1: $-x,2-y,z-\frac{1}{2}$ & \#2: $x+1,y,z$ \\
        \#3: $x,y-1,z$ & \#4: $1-x,2-y,z-\frac{1}{2}$ \\
        \#5: $-x,1-y,z-\frac{1}{2}$ & \#11: $-x,2-y,z+\frac{1}{2}$ \\
    \end{tabular}
    \end{minipage}
\end{table}

% =====================================================
% APPENDIX TRANSPORT FIGURE
% =====================================================

\begin{figure}[!htbp]
    \centering
    \includegraphics[width=0.85\textwidth]
    {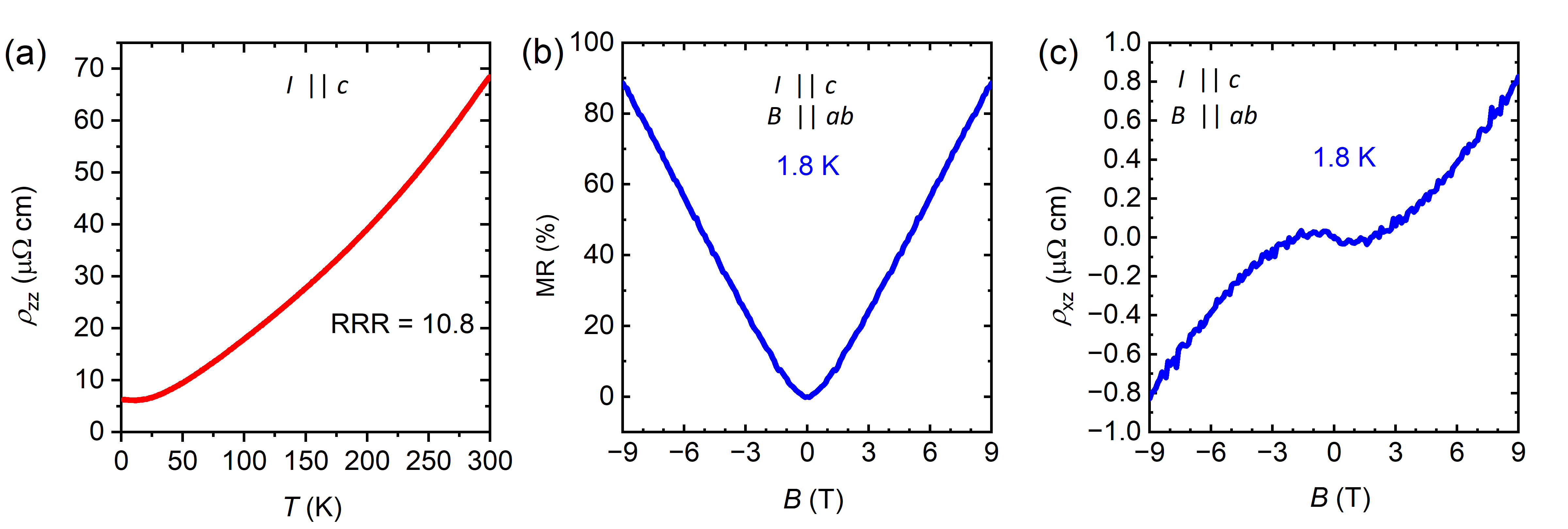}

    \caption{\small
    \textbf{Electrical transport measurements of
    second Sb-flux CrSb sample}:
    (a) Resistivity as a function of temperature with $I\parallel c$ 
    which shows a large RRR value
    (b) Magnetoresistance at 1.8 K
    (c) Hall at 1.8 K along $I\parallel c$ and $B\parallel ab$ orientations.
    }
    \label{TransportSb}
\end{figure}

\begin{figure}[!htbp] 
\centering
\includegraphics[width=0.5\textwidth]{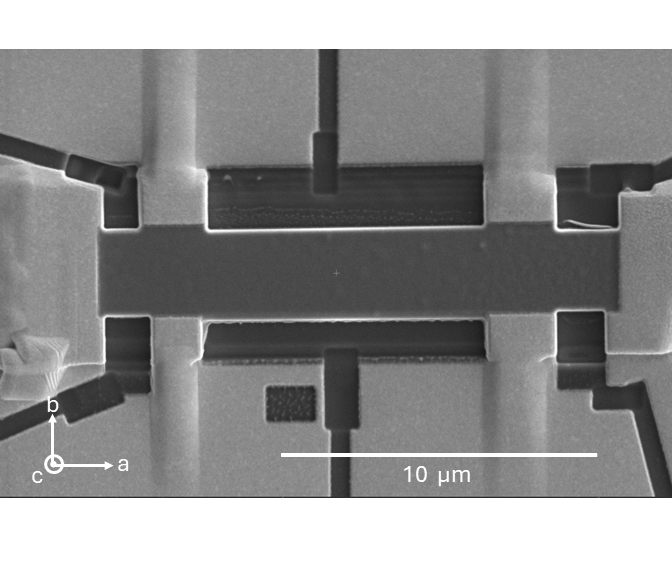} \caption{\small \textbf{Nano-fabricated CrSb device:} a Hall bar device prepared by Pt deposition using a Ga$^{+}$-based FIB for electrical transport measurements of the Bi-flux-grown CrSb sample for the $I \parallel a$ and $B \parallel c$ orientation.} 
\label{device}
\end{figure}

\end{document}